\documentclass[a4paper,12pt]{article}

\usepackage{amsmath, amssymb, amsfonts}
\usepackage{graphicx, color, subfigure}
\usepackage{calc}
\usepackage{amsthm}
\usepackage{natbib}
\usepackage{mathdots, siunitx}
\usepackage{enumitem}
\usepackage{standardmacros}
\usepackage{setspace, authblk}

\title{Intelligent Compaction and Quality Assurance of Roller Measurement Values utilizing Backfitting and Multiresolution Scale Space Analysis}

\author[1]{Daniel K. Heersink\thanks{To whom correspondence should be addressed. Email: daniel.heersink@gmail.com}} \author[1]{Reinhard Furrer}
\author[2]{Mike A. Mooney}
\affil[1]{Institute of Mathematics, University of Zurich, CH-8057 Zurich}
\affil[2]{Colorado School of Mines, Golden, CO 80401, USA}
\date{}

\begin{document}
\maketitle
\doublespacing
\begin{quote}
  Modern earthwork compaction rollers collect location and compaction information as they traverse a compaction site. These roller measurement values present a challenging spatio-temporal statistical problem that requires careful implementation of a proper stochastic model and estimation procedure. \cite{Heer:Furr:13} proposed a sequential, spatial mixed-effects model and a sequential, spatial backfitting routine for estimation of the modeling terms for such data. The estimated fields produced from this backfitting procedure are analyzed using a multiresolution scale space analysis developed by \cite{Holm:etal:11}. This image analysis is proposed as a viable solution to improved intelligent compaction and quality assurance of the compaction process.
\end{quote}

\section{Modern earthwork compaction}\label{sec:intro}
The first modern earthwork compaction rollers designed for continuous compaction control (CCC) were used in practice starting in the 1970s in the European community. CCC is a method of documenting compaction and is used to achieve homogeneous compaction in a minimum time \citep{Thur:Sand:00}. Rudimentary intelligent compaction (IC) technology was first available in the late 1990s. IC is an automated system that adjusts roller operation parameters for optimal compaction based on CCC data \citep{Sche:etal:07}. IC is a development aimed at improving quality assurance (QA) of the compaction process. Utilization of spatial uncertainty in modeling and estimation of this data will lead to improved IC and QA.

Each roller manufacturer has developed a proprietary measurement of soil stiffness used for CCC. This measure of stiffness, coupled with GPS coordinates of the roller when the measurement is taken is termed the roller measurement value (RMV). Current use of RMVs is the identification of potential areas of soft, or weak, spots. Acceptance of these areas is based on the weak spots meeting prespecified criteria \citep{Moon:etal:10}.

\subsection{Roller Measurement Values (RMVs)}
\begin{figure}[h!]
  \centering
  \includegraphics[height=6cm]{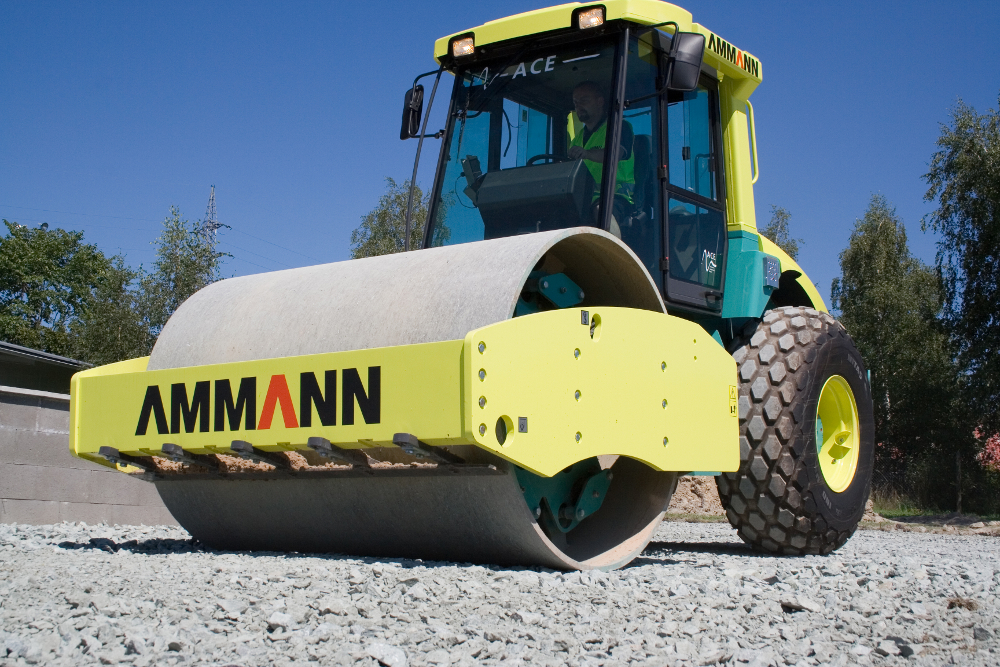}
  \caption{Ammann roller at work.}
  \label{fig:roller}
\end{figure}

A typical smooth drum roller has a diameter of approximately 1m and is approximately 2m long and is outfitted with a sensor and GPS system. The sensor and the GPS system record at set, and not necessarily identical, frequencies. The RMV frequency is the minimum of these two frequencies. Each RMV reflects an aggregated volume of soil measuring a bulb extending to a depth of approximately 1m with a diameter of 0.5--0.6m \citep{Faca:09}. Typical construction practice is to compact in segments of road 10--15m wide and 50--100m long. See Figure \ref{fig:roller} for a representative roller manufactured by Ammann.

\subsection{Minnesota Construction Site} \label{sec:data}
\begin{figure}[h!]
  \centering
  \includegraphics[height=6cm]{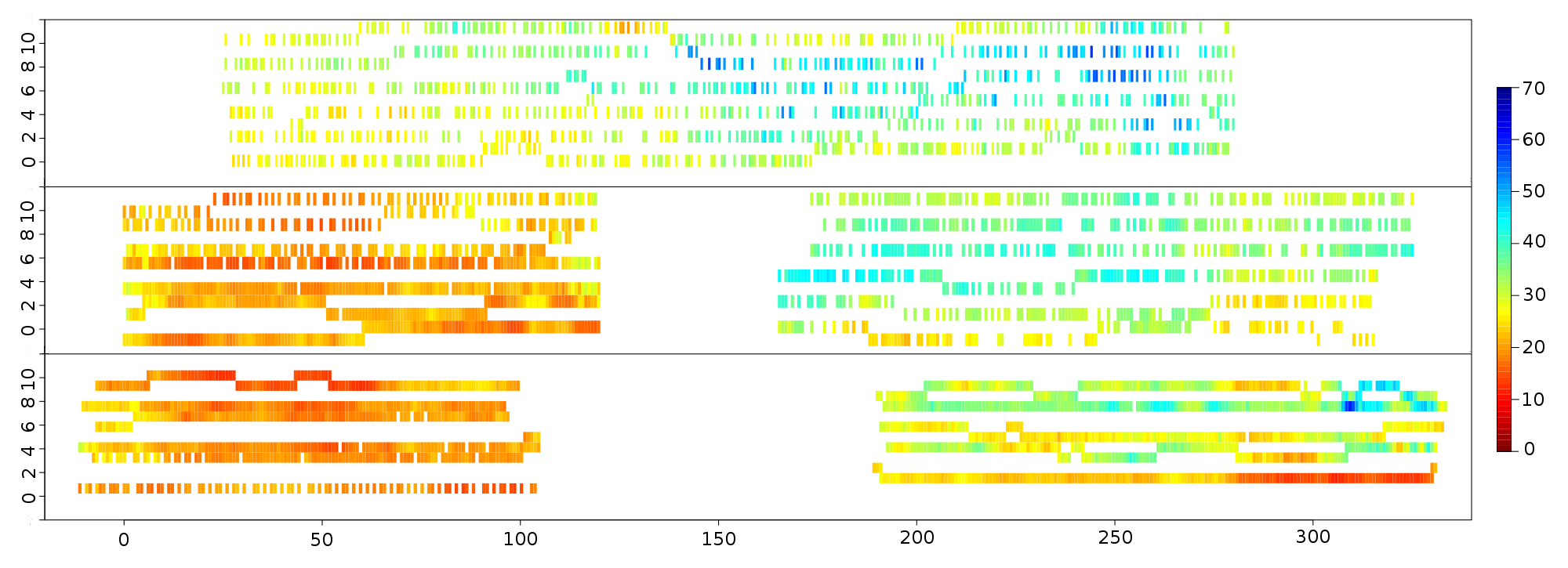}
  \caption{Plot of subsurface (top), subgrade (middle), and base (bottom) data from cells 27 (right) and 28 (left) from the test bed in Albertville, MN, USA. $x$-direction is the direction of driving.}
  \label{fig:data}
\end{figure}

As part of the NCHRP 21-09 project of the Transportation Research Board of The National Academies, Dr. Mike Mooney (Colorado School of Mines, Golden, CO) and his team collected RMV data from a test bed using a smooth, vibrating drum roller manufactured by Ammann \citep{Moon:etal:10}. The test bed lies along a stretch of road adjacent to Interstate 94 (\ang{45;15;45},\,\ang{-93;42;37}). The test bed is approximately 300 meters in length by 15 meters in width and divided into two cells, labeled 27 and 28. Measurements from the roller of the subsurface, subgrade, and base layers of the new road under construction were recorded. Figure \ref{fig:data} is a plot of the data from both cells and all three layers.

Data from each layer was collected separately for each of the two cells, except for the subsurface layer where the roller drove continuously over both cells. Analysis was performed on each cell separately as the standard construction procedure focuses on one cell at a time. The subsurface layer was thus split into two cells at the cell boundary.

The subsurface layer is the existing material (clay) at the construction site. The subgrade layers consist of moisture conditioned clay and the base layers consist of a granular composite material \citep{Moon:etal:10}. The roller traversed the construction site in five to seven lanes, with the coordinates calculated by on-board GPS. Thus, the locations of the observation vector for each layer are unique. 

\section{Quality Assurance and Intelligent Compaction}
The modeling and backfitting estimation routines applied to the Minnesota RMV dataset of \cite{Heer:Furr:13} can be implemented into a more comprehensive program improving quality assurance of the compaction process and also improving intelligent compaction. \cite{Holm:etal:11} developed scale space multiresolution analysis methods for image processing using a Bayesian framework. This methodology is demonstrated using several test images and is applied to climate change prediction fields. The method can also be used to analyze images of RMV estimates to identify weak, or soft, spots and large variations in the compaction area, i.e. QA and IC.

\subsection{Backfitting of RMVs}
The backfitting algorithm has been employed on a wide range of additive models, e.g. \cite{Furr:Sain:09}, \cite{Buja:etal:89}, \cite{Brei:Frie:85}. The classical model for the backfitting algorithm is $\y = \X\bbeta+\balpha+\bvarepsilon$, where $\var(\balpha) = \bSigma(\balpha)$ and $\var(\bvarepsilon) = \sigma^2\I$, $\balpha$ independent of $\bvarepsilon$. If the parameters $\btheta$ and $\sigma$ are known, the backfitting algorithm then iteratively estimates the fixed effects using generalized least squares and the spatial effects using spatial smoothing:
\begin{align*}
  \widehat{\bbeta} &= \left(\X^T\bSigma_\y^{-1}\X\right)^{-1}\X^T\bSigma_\y^{-1}\y \;\text{ (generalized least-squares estimator),} \\ \widehat{\balpha} &= \bSigma(\btheta)\bSigma_\y^{-1}\left(\y - \X\widehat{\bbeta}\right) \;\text{ (spatial smoothing)}, 
\end{align*}
where $\bSigma_\y = \var(\y) = \bSigma(\btheta) + \sigma^2\I$, the covariance matrix of the observations. The spatial backfitting algorithm produces new iterative estimates until they converge, i.e. the estimates no longer differ with each iteration, up to a small number.

\cite{Heer:Furr:13} developed a sequential, spatial mixed-effects model and sequential backfitting methodology for complex spatial structures and apply that methodology to the dataset detailed in Section \ref{sec:data}. To aid in computational time, a spherical covariance structure is assumed as this produces sparse matrices. The model includes a state-space formulation to handle the unique data observation locations. The model assumed is of a spatial ``autoregressive'' type:
\begin{align*}
  \y_1 &= \H_1\X_1\bbeta_1 + \H_1\balpha_1 + \bvarepsilon_1 \\ 
  \y_2 &= \H_2\X_2\bbeta_2 + c\H_2\balpha_1 + \H_2\balpha_2 + \bvarepsilon_2 \\ 
  \y_3 &= \H_3\X_3\bbeta_3 + c^2\H_3\balpha_1 + c\H_3\balpha_2 + \H_3\balpha_3 + \bvarepsilon_3,
\end{align*}
where $c$ represents the level of previously compacted layers the roller ``sees'' while compacting the current topmost layer of material, $\balpha_1, \balpha_2, \balpha_3$ are independent Gaussian random fields, and $\bvarepsilon_1, \bvarepsilon_2, \bvarepsilon_3$ are uncorrelated white noise processes, mutually independent of the $\balpha_i$s. $\X_1, \X_2, \X_3$ are full rank matrices of covariates consisting of an intercept, centered and scaled $(x, y)$-coordinates of the roller measurement, centered and scaled $x$-coordinates of the second and third power, and the driving direction of the roller (1 for right-to-left, 0 for left-to-right). Only higher order powers of the $x$-coordinate were used due to the scale difference in the two directions. Centering and scaling of the coordinates addresses the range anisotropy in RMVs, see \citet{Faca:etal:10} for discussion of anisotropy in RMVs. $\H_1, \H_2, \H_3$ are matrices mapping coordinates of the process level to observed locations.

The process level estimates of the compaction process were obtained from the Sequential Backfitting Algorithm \citep{Heer:Furr:13}. These estimates can then be used as ``images'' of the compaction site for each layer of compaction and each cell of the site.

In practice, there is a site specific threshold value of compaction required for the layer to be deemed sufficiently compacted. This threshold value can be dependent on the current material of compaction. For this analysis, a threshold value of 20 was used for all cells and layers for demonstration. The threshold value was subtracted from all images such that values less than zero represent areas that are too soft. Theoretically, the images are invariant to an additive constant such as this thresholding procedure, i.e. the thresholding does not change the range of the estimates thus the images are produced on the same color scale, irrespective of the threshold value. The thresholding was done for a practical advantage of more easily identifying soft areas.

\subsection{Multiresolution Scale Space Analysis}
To investigate significant features of an image at multiple scales, \cite{Holm:etal:11} developed a Bayesian framework of image processing at multiple resolutions. This multiresolution scale space analysis identifies credible regions of the image that correspond to positive and negative regions. Multiresolution scale space analysis is a method of simultaneously smoothing an input, such as data or an image, at several levels. Each smooth of the input provides a different scale of information.

Multiresolution scale space analysis was applied to the six images obtained from the backfitting algorithm to identify which features in the images are real features (in the Bayesian confidence region sense) and which are artifacts of random variation. The multiresolution scale space analysis uses a Bayesian framework. Each image created from the backfitting algorithm consists of the process level estimates: $\widehat{\y}_t = \X_t\widehat{\bbeta}_t + \sum_k=1^t c^{t-k}\widehat{\balpha}_k$, $t = 1, 2, 3$.

500 samples were then drawn from the ``posterior'' of $\bbeta_1$, $\bbeta_2$, $\bbeta_3$, $\balpha_1$, $\balpha_2$, and $\balpha_3$. Where 
\begin{align*}
  \bbeta_t &\sim \cN\left(\widehat{\bbeta}_t, \left(\X_t\T\H_t\T\bSigma_{y,t}^{-1}\H_t\X_t\right)^{-1}\right) \\
  \balpha_t &\sim \cN\left(\widehat{\balpha}_t, \bSigma_t\H_t\T\bSigma_{y,t}^{-1} \left(\I - \H_t\X_t\left(\X_t\T\H_t\T\bSigma_{y,t}^{-1}\H_t\X_t\right)^{-1}\X_t\T\H_t\T\bSigma_{y,t}^{-1}\right) \H_t\bSigma_t\right),
\end{align*}
and $\bSigma_{y,t} = \H_t\bSigma_t\H_t\T + \sigma_t^2\I$.

\begin{figure}[h!]
  \centering
  \includegraphics[height=8cm]{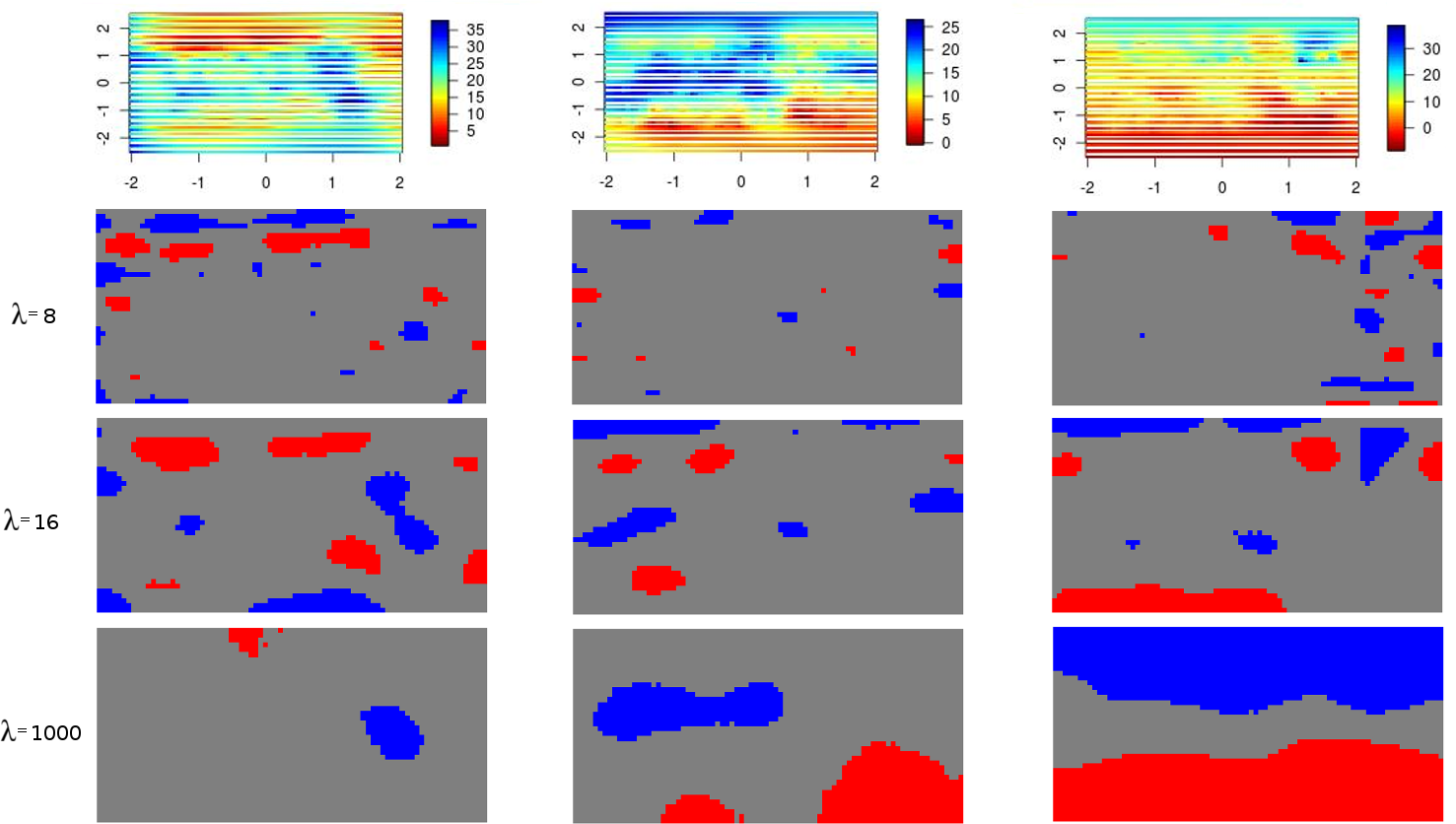}
  \caption{Plot, on a scaled coordinate system, of estimated subsurface (left), subgrade (middle), and base (right) for cell 27 from the test bed in Albertville, MN, USA (top). Credibility plots from a multiresolution scale space analysis are depicted in the bottom three plots for different values of $\lambda$. Red corresponds to softer RMVs. All credibility plots for $\lambda = \infty$ are solid blue, indicating an overall sufficient compaction has been attained with the given threshold.}
  \label{fig:cell27}
\end{figure}

\begin{figure}[h!]
  \centering
  \includegraphics[height=8cm]{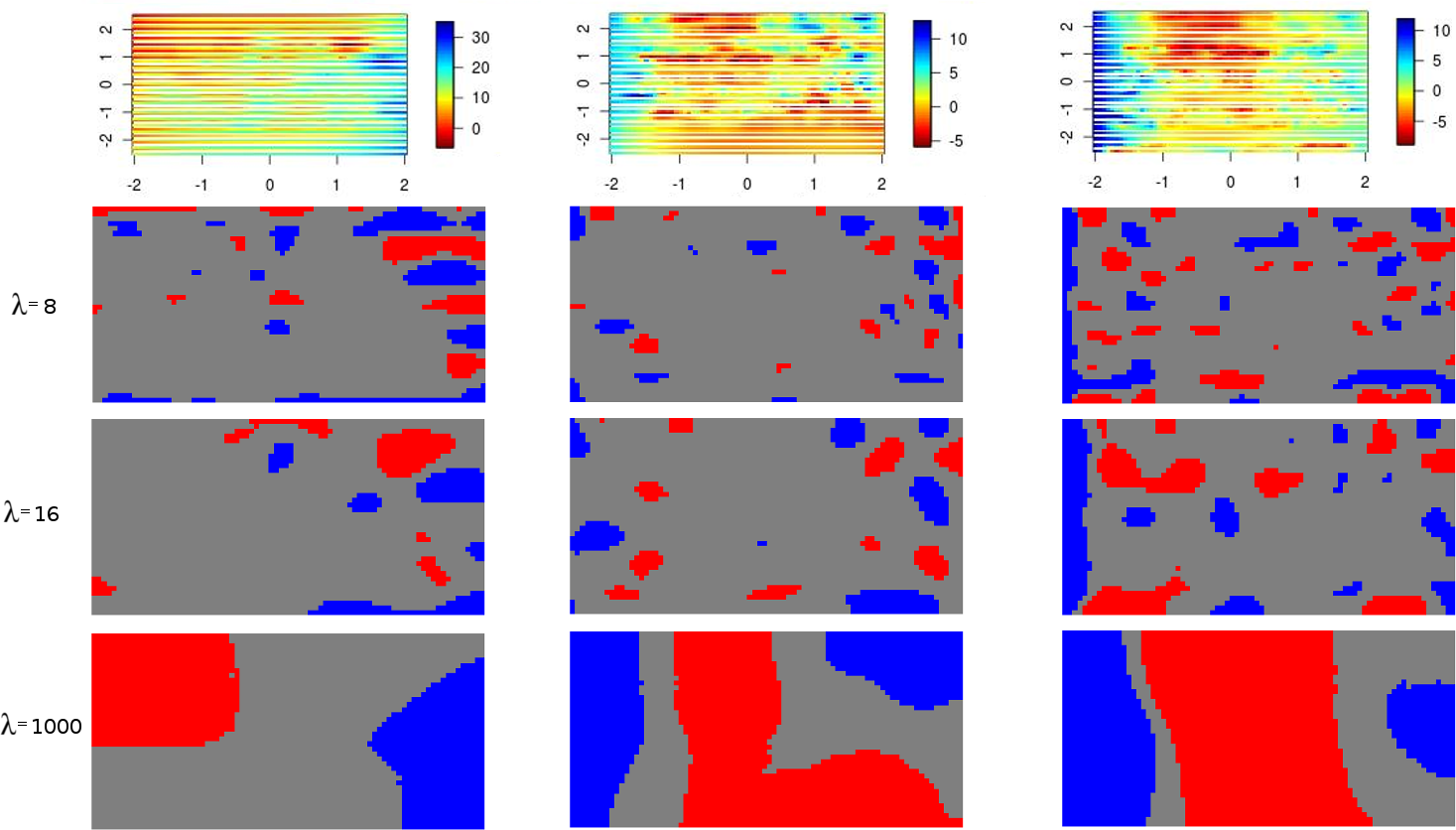}
  \caption{Plot, on a scaled coordinate system, of estimated subsurface (left), subgrade (middle), and base (right) for cell 28 from the test bed in Albertville, MN, USA (top). Credibility plots from a multiresolution scale space analysis are depicted in the bottom three plots for different values of $\lambda$. Red corresponds to softer RMVs. All credibility plots for $\lambda = \infty$ are solid blue, indicating an overall sufficient compaction has been attained with the given threshold.}
  \label{fig:cell28}
\end{figure}

Figure \ref{fig:cell27} depicts the results of this multiresolution scale space analysis for the subsurface, subgrade, and base layers of cell 27. Figure \ref{fig:cell28} depicts the analysis for cell 28. The set of smoothing levels used was [$8, 16, 1000, \infty$]. The smallest smoothing level corresponds to a small scale structure on the order of 5m. The largest smoothing levels correspond to an overall mean and large scale mean structure on the order of 75m. The intermediate smoothing level is smoothing of an intermediate order.

\section{Discussion}
The blue color from the images generated by the multiresolution scale space analysis indicates credible regions of sufficient compaction. The red color indicates credible regions where a sufficient level of compaction is suspect. The $\lambda = 8, 16$ images are most informative for identifying areas requiring more compaction as they identify regions on the same scale as the roller. The larger $\lambda$ values provide a more general idea of the compaction of the entire cell by depicting a gradient of hard to soft values.

Cell 28 is consistently softer than cell 27 through all layers of compaction. This
is an example of heterogeneity in the pre-existing material at the construction site, as represented by the subsurface layer. The features of the subsurface layer are inherited by subsequent layers of compaction. The red spots on the generated images, especially at the $\lambda = 8$ smoothing level, can be seen in all three layers. This is an expected result as each layer of compaction is on the order of 20cm thick and the roller measures to a depth on the order of 1m \citep{Faca:09}. 

Cell 27 displays a gradient of more compact to less compact material from top to bottom in the subgrade layer. This gradient is still evident in the base layer. For cell 28, the a gradient from right to left in the subsurface layer is evident. This gradient changes to a high to low to high gradient moving left to right in the subgrade layer and remains through the base layer.

The figures also detail credible regions of heterogeneity of the compaction region. These areas could have been compacted more to achieve a more homogeneous compaction and improve QA of the construction. The small RMVs found in the base layer of cell 27 could also be identified while compaction is in progress and roller parameters altered to better compact that region of the cell, i.e. IC.

An image of the estimated process level that contained all red and orange, i.e. is everywhere below the threshold value, would return a solid red credibility plot at the highest smoothing levels.

The detailed methodology of sequential, spatial backfitting of RMVs coupled with multiresolution scale space analysis of the resultant estimate images can be utilized to improve IC. By implementing the sequential, spatial mixed-effects model, the spatial uncertainty in RMV data is used to generate estimates of the true compaction level. The use of spherical covariance matrices speeds computation time of the estimation and the resultant images can be produced at a resolution that provides speed in computation of the multiresolution scale space analysis step. Implementation of such a scheme is time effective and could improve the rudimentary IC currently utilized.

The sequential, spatial mixed-effects model uses spatial, random processes in its additive decomposition. Splines can also replace these random processes. The estimation of a spline is mathematically equivalent to the universal kriging done in this paper, as detailed in \citet{Heer:Furr:13}. The literature on splines is extensive and computational feasibility can be maintained, i.e. \cite{Wahb:90}, \cite{Eile:etal:96}, \cite{Marx:Eile:98}, \cite{Eile:Marx:04}.

\bibliographystyle{mywiley}
\bibliography{mybib}
\end{document}